\documentclass[sigplan,10pt]{acmart}

\renewcommand\footnotetextcopyrightpermission[1]{}
\setcopyright{none}
\acmConference{}{}{}
\acmYear{2026}
\copyrightyear{2026}
\acmDOI{}
\acmISBN{}

\usepackage{graphicx}
\usepackage{booktabs}
\usepackage{listings}
\usepackage{amsmath}
\usepackage{multirow}
\usepackage{enumitem}
\usepackage{subcaption}
\usepackage{placeins}
\usepackage{xspace}
\usepackage{pifont}
\usepackage{tikz}
\usetikzlibrary{arrows.meta,positioning,fit,calc}

\makeatletter
\renewcommand\footnotesize{\@setfontsize\footnotesize\@xpt{12}}
\renewcommand\small{\@setfontsize\small\@xpt{12}}
\makeatother
\makeatletter
\AtBeginDocument{%
  \def\ps@plain{\ps@empty
    \def\@oddfoot{\normalfont\normalsize\hfil\thepage\hfil}%
    \let\@evenfoot\@oddfoot}%
  \pagestyle{plain}}
\makeatother

\newcommand{\sys}{ConStore\xspace}
\newcommand{\ccs}{CCS\xspace}
\newcommand{\ccf}{GCF\xspace}
\newcommand{\rpag}{RPAG\xspace}
\newcommand{\wasr}{WASR\xspace}
\newcommand{\sicp}{SICP\xspace}

\begin{document}

\date{}

\title{ConStore: Search-Overlapped Prefetch and \\Co-Retrieval-Aware Content Storage for \\Disk-Based ANNS}

\author{Zhutao Zhuang}
\email{zhuangzht@mail2.sysu.edu.cn}

\author{Nong Xiao}
\email{xiaon6@mail.sysu.edu.cn}

\author{Zhiguang Chen}
\email{chenzhg29@mail.sysu.edu.cn}

\author{Yue Yu}
\email{yuy@pcl.ac.cn}

\makeatletter
\renewcommand{\@mkauthors}{%
  \global\setbox\mktitle@bx=\vbox{%
    \unvbox\mktitle@bx
    \centering
    {\@authorfont
      Zhutao Zhuang\textsuperscript{1,2}\quad
      Nong Xiao\textsuperscript{1}\quad
      Zhiguang Chen\textsuperscript{1}\quad
      Yue Yu\textsuperscript{2}\par}
    \medskip
    {\@affiliationfont
      \textsuperscript{1}Sun Yat-sen University\quad
      \textsuperscript{2}Pengcheng Laboratory\par
      \nolinkurl{zhuangzht@mail2.sysu.edu.cn}\quad
      \nolinkurl{xiaon6@mail.sysu.edu.cn}\par
      \nolinkurl{chenzhg29@mail.sysu.edu.cn}\quad
      \nolinkurl{yuy@pcl.ac.cn}\par}
    \bigskip}}
\makeatother

\begin{abstract}
Disk-based approximate nearest-neighbor search (ANNS) supports
retrieval-augmented generation (RAG), search, recommendation, and advertising
over corpora that exceed memory capacity. Existing ANNS research often
focuses on vector search at small top-$K$, whereas industrial-scale
applications require larger retrieval sets. Fetching the corresponding
content from storage remains comparatively underexplored. Our profiling
across million- to billion-vector corpora shows that, at large $K$, content
fetch latency can match or exceed vector-search latency and cannot be
ignored in end-to-end retrieval.

We present \sys, a content storage engine combining two complementary
 techniques to reduce end-to-end retrieval latency.
{Speculative Inline Content Prefetch} (\sicp) commonly abstracts 
graph-based and cluster-based ANNS as progressive SSD I/O that 
refines an in-flight candidate
queue. It prefetches content for stable leading candidates under a bounded
I/O budget, overlapping content reads with search without changing the final
top-$K$ results.
{Co-Retrieval-Pattern-Aware Grouping} (\rpag) uses query logs to place
frequently co-retrieved documents in contiguous segments.
Workload-Adaptive Segment Reorganization (\wasr) adapts this layout to
workload shifts by selectively regrouping segments under a bounded rewrite
budget.
{Group Content Fetch} (\ccf) exploits this layout to coalesce
per-document random reads into batched, page-aware I/O requests.
\sicp works independently of the content layout, while \rpag and \ccf
reduce fetch cost even without prefetching.

Across three multi-hop QA workloads with $K$ from 10 to 1600,
\sys reduces end-to-end retrieval latency
over the best per-document-read baseline by up to 48\% on DiskANN and 25\%
on SPANN. Separate experiments on 100M- and
billion-vector indexes demonstrate \sicp's applicability at larger scales.
\end{abstract}

\maketitle
\pagestyle{plain}
\thispagestyle{plain}

\section{Introduction}
\label{sec:intro}

Disk-based approximate nearest-neighbor search (ANNS) has been widely adopted
to overcome the memory limitations of large-scale vector search
\cite{subramanya2019diskann,chen2021spann,huang2026helmsman}.
By keeping large indexes on SSD rather than entirely in DRAM, these systems
support retrieval-augmented generation (RAG), search, recommendation, and
advertising over corpora containing millions to billions of vectors
\cite{lewis2020rag,huang2026helmsman}.

Once a query is embedded, the storage-facing ANNS retrieval pipeline has two
phases: \emph{vector search} identifies the top-$K$ item IDs, and
\emph{content fetch} loads their associated payloads from storage for
downstream processing. For example, RAG loads passages for an LLM prompt,
while search and recommendation retrieve item content or features for
re-ranking and presentation. Both phases contribute to end-to-end retrieval
latency.

Prior work has heavily optimized vector search through two main families of
disk-based indexes. Graph-based systems, including DiskANN, PipeANN,
OdinANN, and Starling, optimize graph traversal, storage layout, or update
handling~\cite{subramanya2019diskann,guo2025pipeann,guo2026odinann,wu2024starling}.
Clustering-based systems, including SPANN and FusionANNS, organize vectors
into clusters for efficient candidate retrieval
\cite{chen2021spann,fusionanns2025}.
In contrast, the content-fetch phase that delivers the selected payloads has
received comparatively little attention.

Much ANNS research emphasizes small top-$K$ settings such as $K{=}10$,
but industrial applications retrieve hundreds or thousands of candidates.
Recent study reports search workloads with
$K$ around 100--3{,}000 to supply downstream re-ranking models and advertising
workloads with $K$ up to 3{,}000~\cite{huang2026helmsman}.
These larger retrieval sets challenge the assumption that content fetch is
only a minor overhead. Our profiling across million- to billion-vector
indexes shows that content-fetch latency can match or exceed vector-search
latency at large $K$
(Figures~\ref{fig:motivation-growth} and~\ref{fig:motivation-growth-spann}).
At small $K$, vector search pays an initial cost to locate promising
regions of the index. As $K$ grows, expanding and scoring the candidate set
adds work, while fetching fixed-size payloads requires proportionally more
content reads. In the measured large-$K$ regime, both phases exhibit
approximately linear growth with $K$, making content fetch a first-class
retrieval cost rather than a negligible follow-up.

Conventional retrieval serializes content fetch after vector search, waiting
for the final top-$K$ IDs before issuing content reads. We study graph-based
and clustering-based ANNS and it reveals a shared opportunity: both can be viewed
as \emph{progressive SSD I/O processes that iteratively refine a candidate
queue}. DiskANN reads graph neighborhoods, while SPANN reads and scores
posting lists in batches. In both cases, likely results emerge before search
finishes, allowing some content reads to overlap with the remaining search
I/O.

Choosing when and what to prefetch is non-trivial. Early candidates can
change, wasting content reads; late prefetches leave too little search time
to hide their cost. Variable search tails and competition for SSD bandwidth
therefore require selective, budgeted speculation.

Another challenge is that content is commonly stored in insertion or key
order, unrelated to which items are retrieved together
\cite{wang2021milvus,qdrant2024gridstore,weaviate2024lsmkv}.
Co-retrieved payloads are consequently scattered, requiring many random reads
and limiting opportunities for coalescing.

Embedding-based clustering is an imperfect alternative: proximity in vector
space need not imply co-access. A multi-hop query can retrieve a film page
and its director's biography from different semantic clusters. Grouping
content by observed co-retrieval patterns can capture this locality that
insertion order and embedding clusters miss.

We present \sys, a content storage engine for disk-based ANNS that reduces
end-to-end retrieval latency through two complementary techniques: overlapping
content fetch with search and reducing fetch cost through co-retrieval-aware
layout. Either technique can be deployed independently. Our contributions are:

\begin{enumerate}[leftmargin=*,topsep=2pt,itemsep=1pt]
\item Content-fetch profiling:
Across million- to billion-vector indexes, we show that content-fetch cost
grows approximately linearly with $K$ and can match or exceed vector-search
latency, motivating optimization of the complete search-plus-fetch pipeline.

\item Speculative Inline Content Prefetch (\sicp): 
Using the shared progressive-I/O abstraction, \sicp prefetches content for
stable leading candidates under a bounded I/O budget and fetches missing
payloads after search. It preserves the final top-$K$ and requires only a
document-to-extent lookup, independent of content layout.

\item Co-Retrieval Content Segments (CCS) with Group Content Fetch (GCF): 
Retrieval-Pattern-Aware Grouping (\rpag) uses query logs to group frequently
co-retrieved documents into contiguous \ccs segments.
At query time, \ccf coalesces requested content into batched, page-aware
reads. Together they reduce fetch cost without changing the ANN engine,
with or without \sicp. Workload-Adaptive Segment Reorganization (\wasr)
adapts the layout to workload shifts by selectively regrouping segments
under a bounded rewrite budget when recurring co-retrieval patterns change.

\item Evaluation:
On three multi-hop QA workloads, \sys reduces end-to-end retrieval latency
by up to 48\% on DiskANN and 25\% on SPANN over the best per-document-read
baseline, while preserving retrieval results. Ablations isolate each
technique, and synthetic-payload experiments on 100M- and billion-vector
indexes demonstrate \sicp at larger scales.
\end{enumerate}

\section{Background and Motivation}
\label{sec:motivation}

\subsection{ANNS Retrieval Pipeline}

ANNS-based applications that consume retrieved payloads have two coupled
storage phases: vector search and content fetch. The vector-search phase
probes an ANN index to identify the top-$K$ item IDs; the content-fetch phase
loads the associated text, item content, or features from a content store.
In the conventional pipeline, these reads begin after search returns its
final IDs. Content fetch therefore adds its own storage I/O to end-to-end
retrieval latency rather than being part of the vector-search operation.

\begin{figure}[t]
\centering
\includegraphics[width=\columnwidth]{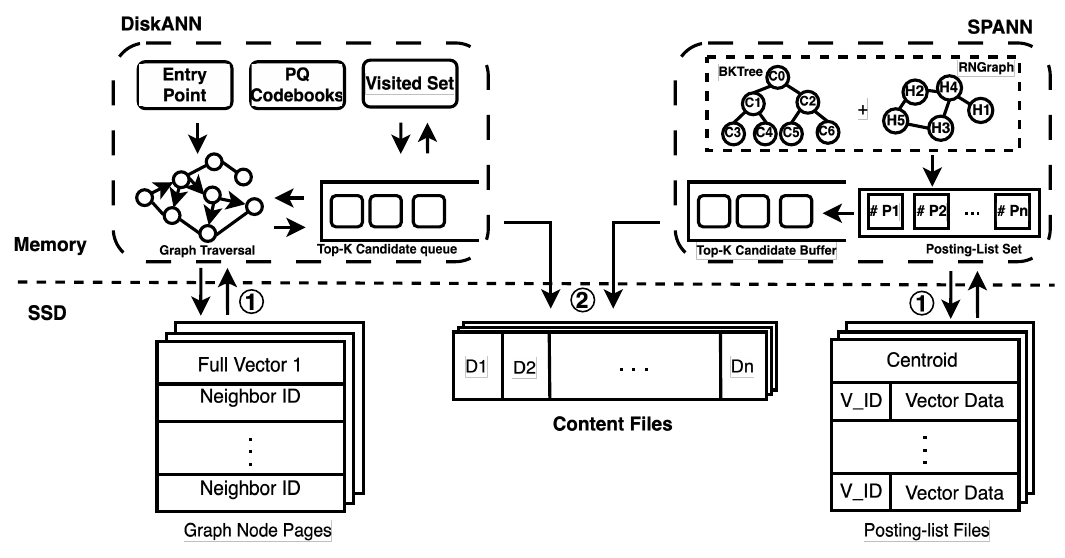}
\caption{Two-phase retrieval over disk-based ANNS, illustrated with a RAG application.}
\label{fig:rag-process}
\end{figure}

ANNS supports more than RAG: semantic search retrieves candidate documents
\cite{karpukhin2020dpr,khattab2020colbert}, recommendation and advertising
retrieve items for downstream ranking or filtering
\cite{huang2026helmsman}, and visual similarity search retrieves related
images~\cite{johnson2019faiss}. In RAG, retrieved passages supply external
knowledge to an LLM~\cite{lewis2020rag,trivedi2023ircot}.

\begin{figure*}[!t]
\centering
\includegraphics{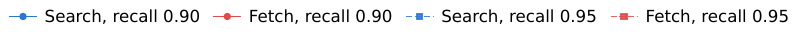}\\[1pt]
\includegraphics[width=0.235\textwidth]{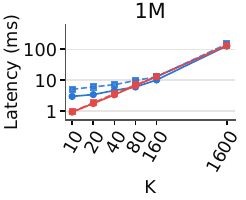}\hfill
\includegraphics[width=0.235\textwidth]{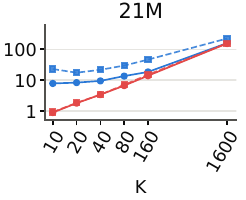}\hfill
\includegraphics[width=0.235\textwidth]{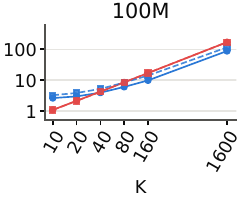}\hfill
\includegraphics[width=0.235\textwidth]{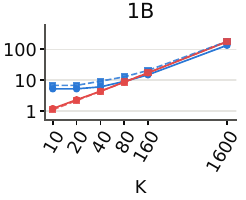}
\caption{Vector search vs.\ content fetch latency growth with $K$ for DiskANN at fixed recall.}
\label{fig:motivation-growth}
\end{figure*}

\begin{figure*}[!t]
\centering
\includegraphics{figures/figure02.pdf}\\[1pt]
\includegraphics[width=0.235\textwidth]{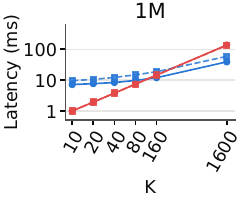}\hfill
\includegraphics[width=0.235\textwidth]{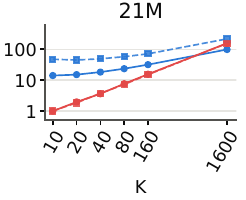}\hfill
\includegraphics[width=0.235\textwidth]{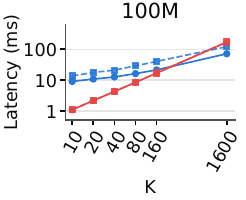}\hfill
\includegraphics[width=0.235\textwidth]{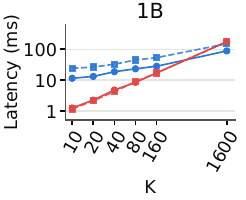}
\caption{Vector search vs.\ content fetch latency growth with $K$ for SPANN.}
\label{fig:motivation-growth-spann}
\end{figure*}

\subsection{Disk-Based ANN Indexes}

Disk-based ANN indexes are used because billion-scale vector corpora cannot
keep all full-precision vectors and search metadata in DRAM at reasonable cost.
Modern disk-based ANN systems mainly fall into two families. The first is
graph-based search, including DiskANN, PipeANN, OdinANN, and
Starling~\cite{subramanya2019diskann,guo2025pipeann,guo2026odinann,wu2024starling}.
DiskANN keeps a compact in-memory structure, such as entry points and
compressed vectors, while storing full-precision vectors and graph
neighborhoods on SSD. Query processing performs a beam search over the graph:
each hop reads a batch of candidate neighbors from SSD, scores the newly loaded
vectors, updates the candidate queue, and issues the next batch of graph reads.
Thus the top-$K$ result is refined over multiple SSD-I/O rounds rather than
being produced by a single blocking lookup.

The second family is clustering-based, IVF-style search, represented by
SPANN and extended by systems such as FusionANNS and
Helmsman~\cite{chen2021spann,fusionanns2025,huang2026helmsman}.
SPANN uses an in-memory head index to route a query
to nearby clusters, while the vectors inside each cluster are stored as posting
lists on SSD. The refinement phase reads selected posting lists progressively,
often one list or a small internal batch at a time, scores their vectors, and
merges them into the current result buffer. Its progress signal is therefore
posting-list completion rather than graph-hop completion.

\subsection{Motivations and Challenges}

Much ANNS research focuses on small top-$K$, such as $K{=}10$, where
content fetch can be a relatively small part of retrieval latency. Industrial
workloads often need larger candidate sets: Helmsman reports search $K$ around 100--3{,}000 
for re-ranking and
advertising $K$ up to 3{,}000~\cite{huang2026helmsman}.
Our profiling shows that content fetch can match or exceed vector-search
latency at large $K$, challenging the assumption that search always
dominates the retrieval pipeline.

Vector-search latency depends on both index size $N$ and result count $K$.
At small $K$, navigation to promising regions can dominate search time.
As $K$ grows, candidate expansion and scoring add work, while fetching
fixed-size payloads requires proportionally more content reads.
Figures~\ref{fig:motivation-growth} and~\ref{fig:motivation-growth-spann}
show this trend from small- to large-scale indexes: in the measured
large-$K$ regime, both phases grow approximately linearly with $K$, making
content fetch non-negligible.

\noindent\textbf{Parallel post-search fetch is not enough.}
Parallel payload reads after search can reduce an isolated query's fetch
latency, but they leave the two phases still serialized
with T\_search + T\_fetch. Under concurrent serving,
one query's burst of content reads competes for SSD bandwidth and queue
capacity with other queries' graph or posting-list reads. Uncontrolled
parallelism can therefore inflate search latency rather than remove the
bottleneck.

\noindent\textbf{Overlap content fetch with vector search.}
Despite their different index structures, DiskANN and SPANN expose the same
opportunity: progressive SSD I/O refines a candidate queue before the final
top-$K$ is known. The key insight is that the head of this queue can stabilize
early, allowing likely result payloads to be prefetched while search
continues rather than waiting for it to finish.

Designing an effective prefetch mechanism is non-trivial.
Prefetching too early wastes bandwidth
because the candidate queue is still unstable; prefetching too late leaves too
little search tail to hide content I/O. The right trigger also depends on
$K$, query difficulty, and the selected ANN backend, because each changes both
candidate stability and remaining search time. A useful mechanism must expose
enough overlap to matter while bounding wasted reads and leaving time for final
top-$K$ verification plus any residual content fetch.

\noindent\textbf{Existing content stores ignore co-retrieval.}
Overlap hides latency but does not reduce the I/O imposed by content layout.
Production vector databases commonly organize payloads by insertion or key
order, not co-retrieval~\cite{qdrant2024gridstore,wang2021milvus,weaviate2024lsmkv}.
Frequently co-retrieved items can therefore remain scattered, requiring
separate reads and limiting coalescing.

Embedding-based clustering is only a proxy for co-retrieval locality.
A question about a film director's birthplace may need both the film page
and the director's biography, even if their embeddings lie in different
clusters. Such documents benefit from co-location based on observed
co-access rather than embedding proximity alone.

\section{\sys Overview}
\label{sec:design}

\sys\ reduces end-to-end retrieval latency by overlapping content reads
with vector search and reducing the I/O required to fetch content.
Prefetching exploits available SSD bandwidth during search under a bounded
I/O budget; layout optimization reduces content-I/O demand.

\begin{figure}[t]
\centering
\includegraphics[width=\columnwidth]{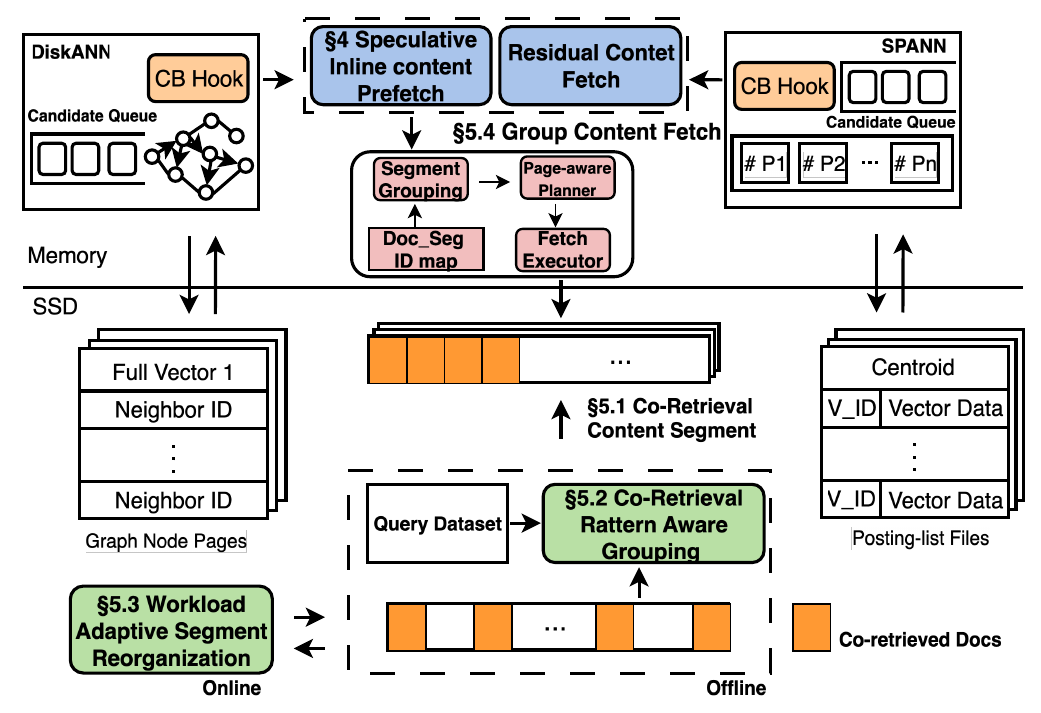}
\caption{\sys architecture. }
\label{fig:vecstore-arch}
\end{figure}

The two techniques are complementary and independently applicable
(Figure~\ref{fig:vecstore-arch}). First, \sicp\ (\S\ref{sec:ssip})
prefetches likely result payloads from the evolving ANN candidate queue,
overlapping reads with search while preserving the final top-$K$.
It works with any content layout that provides a document-to-extent lookup.

Second, co-retrieval-aware layout with grouped fetch
(\S\ref{sec:layout-section}) reduces per-fetch cost.
\rpag\ groups frequently co-retrieved documents into \ccs\ segments,
and \ccf\ coalesces requested content into page-aware reads.
Workload-Adaptive Segment Reorganization (\wasr, \S\ref{sec:wasr})
adapts this layout to workload shifts by selectively regrouping segments
under a bounded rewrite budget when recurring co-retrieval patterns change.
This path requires no ANN-engine changes and benefits both prefetching
and post-search fetch.

Table~\ref{tab:design-axes} separates \emph{when} reads execute from
\emph{how much I/O} they require. Combined, \sicp\ issues \ccf's grouped
reads during search, sharing the same planner with post-search fetch.

\begin{table}[t]
\centering

\caption{The two techniques of \sys and their primitives}
\label{tab:design-axes}
\begin{tabular}{lll}
\toprule
 & \textbf{Wall-clock time} & \textbf{Per-fetch cost} \\
\midrule
\textbf{Runtime path}   & \sicp (\S\ref{sec:ssip}) & \ccf\ (\S\ref{sec:ccf}) \\
\textbf{Storage layout} & --- & \ccs + \rpag \\
                        &  & (\S\ref{sec:ccs}, \S\ref{sec:layout}) \\
\bottomrule
\end{tabular}
\end{table}

\paragraph{Lifetime of a query.}
While the foreground ANN search runs, one background worker observes
read-only candidate snapshots and prefetches through \ccf\ when stability
and budget checks pass. Search runs to completion with its final top-$K$
unchanged. \sys\ then joins the worker and fetches any missing result
payloads through the same \ccf\ path.

\S\ref{sec:ssip} first presents \sicp, the more general of the two
techniques; \S\ref{sec:layout-section} then formalizes the storage and
layout primitives that make each fetched window cheaper.

\section{Speculative Inline Content Prefetch}
\label{sec:ssip}

Prefetching must balance early candidate churn against a shrinking search
tail. \sicp exploits the key insight that leading candidates can stabilize
before the full search completes: it checks read-only \emph{candidate
snapshots}, admits a bounded set of likely results, and overlaps their
page-aware content reads with search. Repeated small prefetches adapt to
later candidates without committing the entire budget to an early snapshot.
A post-search \emph{residual fetch} serves any final top-$K$ documents not
covered by speculation, preserving the search results and recall.
The mechanism requires only a document-to-extent lookup, so it works with
both insertion-order payload files and the layout of
\S\ref{sec:layout-section}.

\S\ref{sec:ssip-design} presents the overall design, \S\ref{sec:ssip-trigger}
develops the policy that decides when each prefetch starts and which
content windows it admits, and \S\ref{sec:ssip-impl} describes the engine-level
hooks that expose search progress.

\subsection{Overall Design}
\label{sec:ssip-design}

\begin{figure}[t]
\centering
\includegraphics[width=\columnwidth]{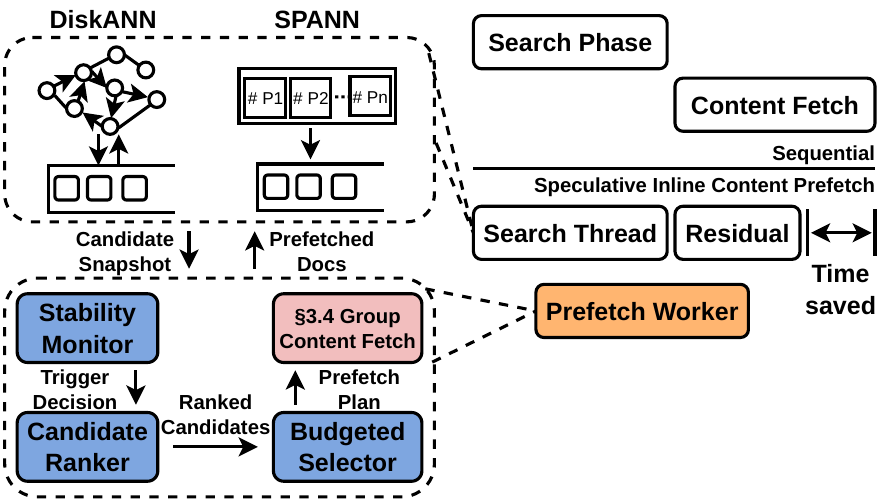}
\caption{Speculative Inline Content Prefetch (\sicp) timeline.}
\label{fig:sicp-timeline}
\end{figure}

Figure~\ref{fig:sicp-timeline} shows the \sicp
process and contrasts it with the sequential baseline. \sicp performs one full
ANN search, but instruments the existing DiskANN/SPANN search loop with a
lightweight read-only progress callback. Each callback reports how far the
search has advanced and copies the current ranked candidate state into a
candidate snapshot.
For DiskANN, progress is measured in graph-search hops; for SPANN, progress is
measured by SSD posting-list entries processed during refinement. A background
\emph{prefetch worker} consumes these snapshots and issues content reads; it
cannot modify, prune, or reorder the foreground vector search.

Before issuing a prefetch, the worker applies the stability and progress checks
of \S\ref{sec:ssip-trigger} to avoid spending bandwidth on transient
candidates. It admits only a bounded number of new candidates, skipping
documents covered by earlier prefetches, so one early snapshot cannot consume
the entire overlap window.
To reduce I/O requests, a page-aware planner maps selected document IDs to
content extents and coalesces nearby extents into page-aligned read windows.
In \sys this planner is \ccf\ (\S\ref{sec:ccf});
over an insertion-order payload file, where co-retrieved documents rarely
share pages, it degenerates to one aligned window per document. The worker records scheduled documents and windows so
later prefetches and the residual fetch do not repeat the same work.

\begin{figure}[t]
\centering
\includegraphics[width=\columnwidth]{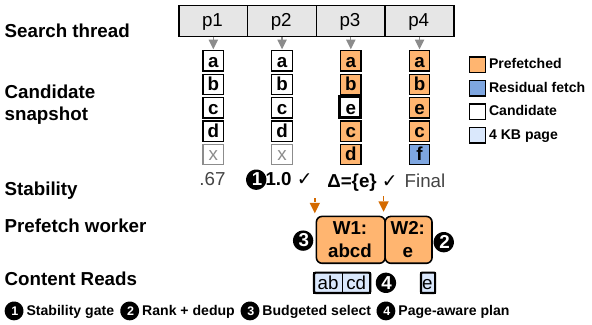}
\caption{\sicp example for one query. }

\label{fig:sicp-example}
\end{figure}

The foreground search is unchanged and continues to its natural completion. It
does not observe prefetch decisions, so it returns the same final top-$K$ as a
non-instrumented run.
After search returns, \sicp computes the residual demand:
final top-$K$ documents that are not already covered by prefetch. This residual
fetch can run while the prefetch worker finishes outstanding work, and it avoids
documents already scheduled by prefetch so the same content is not fetched
twice. If speculation misses, the remaining final documents are fetched normally.
Figure~\ref{fig:sicp-example} walks through these steps for one query.

\noindent\textbf{Bounded I/O overhead.} \sicp bounds speculative work at three levels:
the number of prefetches, the maximum documents admitted per prefetch, and the
per-prefetch I/O budget. It also tracks what previous prefetches have already
selected, so a candidate that remains near the head of the queue is not planned
repeatedly.
The extra I/O beyond the sequential baseline is limited to windows for
speculative candidates that do not survive into the final top-$K$; final
documents that were not prefetched are still served by the residual path.

For one prefetch, the ideal wall-clock saving is bounded by
$\min(T_\text{prefetch\_io},\, T_\text{search\_tail})$, where
\emph{search\_tail} is the remaining vector-search time after the prefetch is
issued and \emph{prefetch\_io} is the worker's content-I/O time for that
prefetch. With multiple prefetches, the same accounting applies repeatedly:
content I/O completed before the search returns is removed from the foreground
fetch path, while any worker wait or residual fetch remains visible.

\subsection{Prefetching Strategy}
\label{sec:ssip-trigger}

The strategy answers two questions continuously throughout the search:
\emph{when} should the next prefetch start, and \emph{what} additional content
should it fetch? A simple
all-or-nothing trigger --- fire once and prefetch the whole current top-$K'$ ---
is brittle. It overcommits when the remaining search tail is short, and it
cannot react when later callbacks promote new documents into the head of the
queue. \sicp instead treats the search tail as a series of small budgets. Each
prefetch spends only a bounded amount of I/O, deduplicates against previous
prefetches, and leaves all misses to the residual fetch.
The strategy has three components.

\noindent\textbf{(1) Progress and stability checks --- when to fire.}
These checks reduce wasted reads from early candidate churn while retaining
time to overlap useful I/O. At every
progress callback, the prefetch worker considers issuing a prefetch only if the
callback is new and the per-query prefetch cap has not been reached. The first
check is progress: DiskANN must have completed enough hops, and SPANN must have
processed enough SSD posting-list entries. The second check is candidate
stability.
\sicp compares the ranked top-$K'$ document set against the previous check using
set-overlap similarity. After the head remains stable for a
configured number of checks, the worker may fire.
A late-progress threshold prevents pathological waiting when the search tail is
almost gone. Later prefetches reuse the same checks but also require a minimum
number of newly promoted documents, so \sicp does not issue a prefetch for a tiny
or duplicate delta.

\noindent\textbf{(2) Budgeted page-aware prefetch --- what to fetch.}
Budgeting limits speculative device pressure, while coalescing reduces the
requests needed to fetch selected candidates. When
prefetching, the prefetch worker ranks the current candidate snapshot and forms
a bounded pool: at most $K\cdot\alpha$ documents, further capped by a
per-prefetch maximum.
Documents already selected by earlier prefetches are excluded. The remaining
candidates are passed to a budgeted selector, which chooses a prefix under the
prefetch I/O budget.
The page-aware planner (\ccf\ in \sys, \S\ref{sec:ccf}) then translates
those selected documents into content extents and merges nearby extents into
page-aligned read windows.

The first prefetch can size its budget from the estimated remaining search tail.
Subsequent prefetches use a small per-prefetch budget so that the worker
continues to track the queue as it evolves, rather than spending the entire
budget on one early snapshot.
\sicp keeps a compact coverage record for scheduled and completed content
windows, so later prefetches avoid both document-level duplicates and page-level
duplicates caused by adjacent documents sharing the same content page.

\noindent\textbf{(3) Residual content fetch --- handling misses.}
The residual path ensures complete final results even when speculation
misses, without duplicating pending reads. After vector
search returns, the final top-$K$ is authoritative. \sicp subtracts all documents
covered by completed prefetch reads.
It also treats documents scheduled by an in-flight prefetch as pending, so the
foreground path does not immediately duplicate the same content work. The
foreground then fetches the remaining final documents with the same page-aware
planner while the prefetch worker finishes outstanding work.

\subsection{Per-Engine Instrumentation}
\label{sec:ssip-impl}

Each engine provides a read-only callback that publishes candidate snapshots
and a progress counter to the shared prefetch worker. The callback leaves
search logic unchanged and requires no suspension or restart.

\noindent\textbf{DiskANN.} On-disk beam search advances through graph-search
hops, reading candidate neighborhoods from SSD and refining the candidate
queue. An optional callback at configurable hop intervals publishes the
current queue and hop counter. The background worker uses successive
snapshots to check stability and select bounded content reads, allowing
multiple prefetches as candidates evolve. The callback does not modify
the queue or traversal; the foreground search continues to completion.

\noindent\textbf{SPANN.} An in-memory head search selects candidate posting
lists, which are read from SSD and scored to refine the result buffer.
We instrument this refinement loop to publish capped, sorted candidate
snapshots at configurable posting-count intervals. The processed-posting
count supplies the progress signal in place of DiskANN's hop counter,
letting the same worker apply stability checks and budgeted prefetching.
Head-index navigation, posting-list selection, and foreground refinement
remain unchanged, preserving the final top-$K$.

\section{Co-Retrieval-Aware Content Layout and Group Fetch}
\label{sec:layout-section}

\sicp (\S\ref{sec:ssip}) hides content I/O behind the search tail but
does not change how much I/O each fetch costs: with an insertion-order
payload store, every prefetched or residual document is still a
separate random read. This section presents the second, independent
technique of \sys, which reduces the per-fetch cost itself. It
consists of an on-disk segment format (\ccs, \S\ref{sec:ccs}), an
offline co-retrieval-aware layout policy (\rpag, \S\ref{sec:layout}),
and an online grouped, page-aware fetch path (\ccf, \S\ref{sec:ccf}).
The three are used by the post-search fetch path on their own, and by
\sicp's prefetch worker when the two techniques are combined.

\subsection{Co-Retrieval Content Segments (\ccs)}
\label{sec:ccs}

A \emph{Co-Retrieval Content Segment} (\ccs) stores a bounded group of
documents contiguously in the order chosen by the layout policy.
Grouping by co-retrieval rather than embedding proximity lets jointly
requested documents share content reads, reducing random I/O even when
their embeddings are far apart.
The offset-addressed format comprises a fixed-size metadata header,
a per-document index of $(\mathit{offset}, \mathit{length})$ pairs,
and a contiguous content block. Embeddings, graph edges, and posting
lists remain in the ANN engine's index, keeping content placement
independent of the search backend.

Offsets support direct lookup of individual payloads. When requested
documents share a segment, \ccf\ (\S\ref{sec:ccf}) reads their index
slab once and coalesces their extents into page-aware windows.
Both post-search fetch and \sicp\ use this path to exploit the locality
created by \rpag\ (\S\ref{sec:layout}).

\sys\ maintains a single
\texttt{doc\_map} that resolves any document ID to its segment and
local index in $O(1)$. The runtime representation is a packed pair of
32-bit integers, so the map costs 8\,B per document, roughly one
percent of the embedding-side ANN index memory in our workloads. This
small fixed metadata cost keeps the fetch path independent of the ANN
engine while avoiding a per-query hash lookup or external metadata
lookup on the critical path.

\subsection{Co-Retrieval-Pattern-Aware Grouping (\rpag)}
\label{sec:layout}

The effectiveness of \ccs\ depends critically on which documents
share a segment. Documents grouped together pay one batched
page-aware read plan per top-$K$ overlap; documents in different
segments require separate segment-level plans. The layout policy is
therefore the lever that controls the per-fetch-cost axis of
Table~\ref{tab:design-axes}.

\begin{figure}[t]
\centering
\includegraphics[width=\columnwidth]{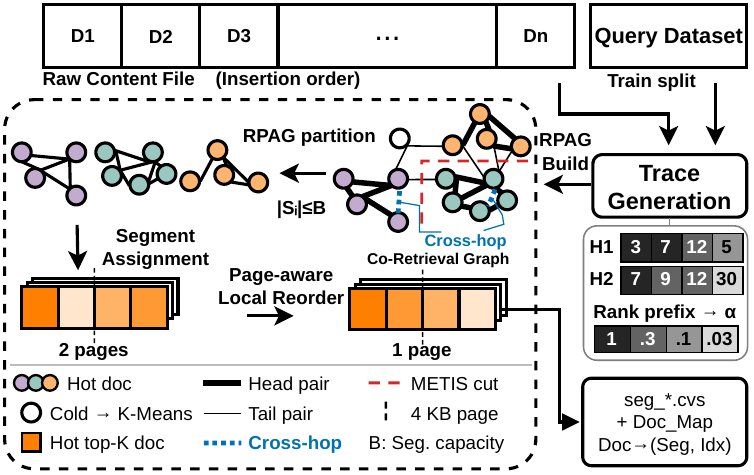}
\caption{\rpag\ layout pipeline. }
\label{fig:rpag-layout}
\end{figure}

\noindent\textbf{Co-retrieval graph and partition objective.}
\rpag\ treats the training retrieval log as a weak supervision signal
for physical locality. Each document is a vertex in a graph
$G=(V,E)$. For a query $q$, let $R_q$ be the ANN result list and
$\rho_q(d)$ be the rank of document $d$ in that list. For every pair
of documents that co-occur in $R_q$, \rpag\ adds an undirected edge
whose contribution depends on the worse rank of the pair. For wider
training lists, \rpag\ can avoid treating all top-$K$ co-occurrences
equally by using prefix-weighted edges: pairs that co-occur in the
head of the result list receive the largest weight, and pairs that
co-occur only in deeper prefixes receive progressively smaller weight.
RPAG instantiates this rule with an ordered prefix set
$\mathcal{C}=\{\kappa_1 < \cdots < \kappa_L=K\}$ and monotonically
decreasing weights $\alpha_1 > \cdots > \alpha_L$:
\[
\begin{aligned}
w_{uv} &= \sum_{q:\,u,v\in R_q} \alpha_{\ell(q,u,v)}, \\
\ell(q,u,v) &= \min\{\ell:
\max(\rho_q(u),\rho_q(v))\leq \kappa_\ell\}.
\end{aligned}
\]
This weighting makes the segment layout reflect the documents most
likely to be requested together by the retriever, while still retaining
weaker evidence from the tail of a wider result list. For multi-round
retrieval traces, edges are added both within a round and across
adjacent rounds, capturing the locality needed by iterative RAG
queries.

The partitioning goal is to group documents so that high-weight edges
stay inside a bounded \ccs\ segment:
\[
\max_{\{S_i\}} \sum_i \sum_{(u,v)\in E,\,u,v\in S_i} w_{uv}
\quad\text{s.t.}\quad |S_i|\leq B.
\]
This is balanced graph partitioning~\cite{karypis1998metis} and
admits direct off-the-shelf solvers. \rpag\ partitions each connected
component of the co-retrieval graph; small components become segments
directly, while larger components are split by METIS into roughly
equal-sized parts. The result is a document-to-segment assignment that
the \ccs\ writer materializes as segment files and a packed
\texttt{doc\_map}. Because the objective mentions only document IDs
and query co-occurrence, it is independent of the ANN engine that will
later produce those IDs.

A pure co-retrieval graph covers only documents observed in the
training log. Treating unobserved documents as arbitrary isolated
vertices would make their placement unstable and would waste segment
capacity on weak evidence. \sys\ therefore uses a hybrid assignment:
documents with sufficient co-retrieval support are placed by \rpag,
while cold or unseen documents inherit an embedding k-means placement.
This fallback is not a separate online path; it is resolved once when
the \ccs\ files are written, so every query still uses the same
constant-time doc-id-to-segment lookup.

\textbf{Page-aware intra-segment refinement.}
Segment membership decides which documents can be fetched together;
the order inside a segment decides whether those documents share the
same SSD pages. We therefore apply a page-aware refinement, RPAG,
after the hybrid \rpag/k-means assignment. RPAG preserves each
segment's document set and constructs a second co-retrieval graph only
among documents already assigned to that segment. It uses the same
prefix-weighted evidence as above inside each existing segment, but now
the objective is linear ordering rather than partitioning:
co-retrieved documents should appear near each other in the segment's
content block so that \ccf\ can cover them with one or a few
page-aligned read windows.

For each touched segment, RPAG greedily packs the intra-segment
graph. It starts from the document with the largest weighted degree,
then repeatedly appends the strongest unplaced neighbor of the current
document; when a connected component is exhausted, it resumes from the
highest-degree remaining document. The \ccs\ writer then rewrites the
segment's content bytes and local id order according to this
permutation, leaving untouched segments byte-identical. This step does
not change recall or segment membership: it only converts the locality
that \rpag\ exposes at segment granularity into page locality that the
page-aware \ccf\ planner can consume.

\subsection{Workload-Adaptive Segment Reorganization (\wasr)}
\label{sec:wasr}

\rpag's layout can become stale when queries begin retrieving different
documents together. We propose \emph{Workload-Adaptive Segment Reorganization}
(\wasr) to adapt the layout to these workload shifts. A background controller
regroups selected segments only when recurring co-retrieval patterns reveal
degraded locality and the expected I/O savings justify rewriting them.
Queries continue using the constant-time \texttt{doc\_map} of their current
layout version while reorganization runs.

\noindent\textbf{Learning recurring locality.}
Real workloads often exhibit skewed popularity, with popular queries and
entities revisited. A learned layout benefits only when document pairs are
retrieved together again. \wasr\ measures \emph{pair recurrence} $r_t$:
the fraction of epoch $t$'s co-retrieval edge weight contributed by pairs
seen in earlier epochs. It retains k-means placement without confirmed
co-retrieval evidence and avoids rewriting for unpredictable workloads.

\noindent\textbf{Online co-retrieval monitor.}
The monitor identifies recurring pairs while aging out obsolete evidence.
It samples completed queries' \emph{final} top-$K$ lists, not provisional
ANN candidates, and aggregates each epoch of $E$ queries into a time bucket.
For documents $u$ and $v$, their recent edge weight is
\[
w_t(u,v)=\sum_{a=0}^{W-1}\gamma^a c_{t-a}(u,v),
\]
where $t$ is the current epoch, $W$ is the number of retained buckets,
$a$ is a bucket's age, $0<\gamma\leq1$ is the decay factor, and
$c_{t-a}(u,v)$ is that epoch's pair weight using \rpag's rank-prefix
weighting (\S\ref{sec:layout}). Bucket rotation expires old evidence.
The monitor also keeps the undecayed occurrence count $n_t(u,v)$ over
these buckets; a pair is \emph{confirmed} when $n_t(u,v)\geq\theta$,
where $\theta$ is the minimum repeat count. This filters one-off
co-occurrences before they can trigger rewrites.

\noindent\textbf{Layout-quality signal and trigger.}
To avoid rewriting for frequency changes that do not harm locality, \wasr\
tracks the \emph{confirmed cut ratio} of assignment $P$, where $P(u)$ is
document $u$'s segment:
\[
\chi_t(P)=\frac{\sum_{(u,v):\,n_t\ge\theta,\;P(u)\ne P(v)} w_t(u,v)}
              {\sum_{(u,v):\,n_t\ge\theta} w_t(u,v)},
\]
the fraction of confirmed edge weight split across segments. With no
confirmed edges, there is no evidence to reorganize. Otherwise, \wasr\
compares $\chi_t$ with the reference $\chi^\ast$ recorded at layout
publication. Planning requires a rise greater than $\delta$ for $h$
consecutive epochs and recurrence $r_t\geq r_{\min}$. Here $\delta$ is
the tolerated degradation, $h$ filters transient shifts, and $r_{\min}$
requires repeated evidence. A plan must also clear the gain threshold below,
avoiding rewrites with little expected benefit.

\noindent\textbf{Budgeted bundle planning.}
\wasr\ reorganizes bundles of whole segments to preserve balanced groups
and page locality without leaving partially dead segments. The planner
charges each confirmed cross-segment edge to both endpoints' segments,
ranks segments by cut weight per byte, and selects them with their heaviest
cut partners under a byte budget. Joint selection lets newly co-retrieved
documents be regrouped. METIS repartitions the bundle into segments of at
most $B$ documents, followed by \rpag's intra-segment ordering; documents
without recent edges retain their previous grouping, packed first-fit.
A plan is accepted only if the recovered cut weight is at least a fraction
$g_{\min}$ of total confirmed weight.

A token bucket adds $b$ times the corpus bytes per epoch and caps accumulated
credit, bounding rewrite volume and bursts after quiet periods. Foreground
queue-depth and tail-latency guardrails further limit the active rewrite rate.

\noindent\textbf{Immutable generations and reclamation.}
To keep queries consistent during reorganization, each layout generation
contains segment files and its own \texttt{doc\_map}. Publication writes new
segments under fresh IDs, hard-links untouched files, records a manifest,
fsyncs, and atomically renames the staging directory. Each query pins one
generation, so it never mixes old and new layouts; the ANN index and recall
remain unchanged. Retired segments are deleted only after the last reader
of the old generation exits. Replacing whole segments avoids holes and
garbage collection within partially live files.

\subsection{Group Content Fetch (\ccf)}
\label{sec:ccf}

\ccf\ is the runtime mechanism that consumes the locality created by
\ccs\ and \rpag. For each query, \sys\ uses the document-to-segment
map and per-document content index described in \S\ref{sec:ccs} to
collect requested documents that fall in the same \ccs\ segment. The
segment is the planning scope, but the read unit is a page-aware
content window rather than the whole segment.

Within a segment, \ccf\ turns the requested documents into a small
window-planning problem. It first orders the requested local document
IDs, reads the corresponding contiguous content-index slab, and
converts each $(\mathit{offset}, \mathit{length})$ entry into an
absolute file extent. Alignment is performed on these absolute offsets,
so the planned windows correspond to the device pages actually read
rather than to offsets relative to the segment's content block.

To reduce requests without excessive overfetch, \ccf\ scans extents in
offset order and greedily merges their page-aligned windows. A merge
that adds no pages is always accepted; otherwise it must satisfy caps
on window size, read amplification (aligned bytes divided by useful
bytes), and the gap between extents. If a cap fails, the planner starts
a new window; a single large document still receives a complete aligned
window. Nearby requested content thus shares I/O, while distant
content is fetched separately rather than reading the whole segment.

The same window plan is used by both \sicp{}'s prefetch worker and the
post-search residual fetch.  After a planned window is issued, documents
whose content ranges are fully covered by fetched pages are marked warm,
allowing the residual path to skip them and fetch only uncovered final
top-$K$ documents. This bookkeeping does not change the ANN result or
the window-building rule; it only shares page coverage between the
speculative and residual uses of \ccf.

The result is fewer content read requests and fewer device pages per
top-$K$ fetch. A \ccf\ hit amortizes the segment's content-index access
and serves multiple co-retrieved documents from one shared window; a
miss simply degenerates to the normal single-document window. This
lowers IOPS demand and queueing pressure without changing ANN search
results or depending on a particular ANN backend.

\section{Evaluation}
\label{sec:eval}

\begin{figure*}[t!]
\centering
\includegraphics{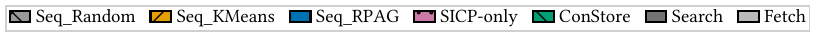}\\[2pt]
\begin{subfigure}[t]{0.32\textwidth}
  \centering
  \includegraphics[width=\linewidth]{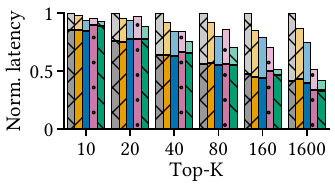}
  \caption{2Wiki.}
\end{subfigure}
\hfill
\begin{subfigure}[t]{0.32\textwidth}
  \centering
  \includegraphics[width=\linewidth]{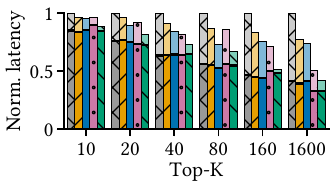}
  \caption{HotpotQA.}
\end{subfigure}
\hfill
\begin{subfigure}[t]{0.32\textwidth}
  \centering
  \includegraphics[width=\linewidth]{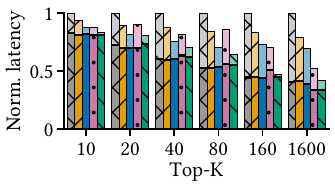}
  \caption{MuSiQue.}
\end{subfigure}
\caption{DiskANN end-to-end latency breakdown by Top-$K$ and dataset. }
\label{fig:e2e-diskann-stacked}
\end{figure*}

\begin{figure*}[!t]
\centering
\includegraphics{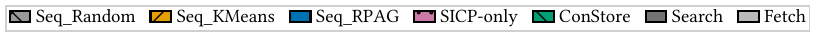}\\[2pt]
\begin{subfigure}[t]{0.32\textwidth}
  \centering
  \includegraphics[width=\linewidth]{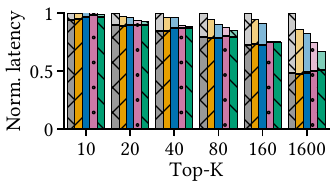}
  \caption{2Wiki.}
\end{subfigure}
\hfill
\begin{subfigure}[t]{0.32\textwidth}
  \centering
  \includegraphics[width=\linewidth]{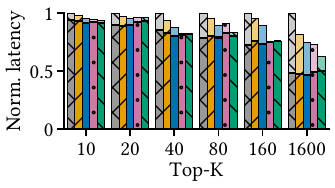}
  \caption{HotpotQA.}
\end{subfigure}
\hfill
\begin{subfigure}[t]{0.32\textwidth}
  \centering
  \includegraphics[width=\linewidth]{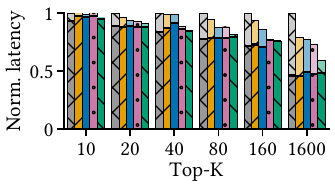}
  \caption{MuSiQue.}
\end{subfigure}
\caption{SPANN end-to-end latency breakdown by Top-$K$ and dataset.}
\label{fig:e2e-spann-stacked}
\end{figure*}

\subsection{Setup}

Experiments run on a two-socket Intel Xeon Gold 6230N server
with 2.30\,GHz 40 cores, 251\,GiB DRAM and a
1.6\,TB Intel SSDPEDME016T4 NVMe SSD.

\noindent\textbf{Datasets.}
The primary corpus is Wikipedia DPR~\cite{karpukhin2020dpr}, with 21M
passages averaging 1\,KB and 768-dimensional embeddings. We use three
multi-hop QA workloads: 2WikiMultihopQA~\cite{ho2020wikimultihop},
HotpotQA~\cite{yang2018hotpotqa}, and MuSiQue~\cite{trivedi2022musique},
containing 12{,}576, 7{,}405, and 2{,}417 questions, respectively.
Each measured question comprises two retrieval hops; we report paired
hop1+hop2 latency. Scaling experiments use SIFT100M and SIFT1B from
BIGANN~\cite{jegou2011bigann}, with synthetic 1\,KB payloads.

\noindent\textbf{RPAG layout construction.} For \ccs segments built with \rpag, we
use a K-prefix co-retrieval objective that
uses cumulative top-$K$ prefixes at $K{\in}\{10,20,40,80\}$ with weights
$1.0$, $0.3$, $0.1$, and $0.03$, respectively. This gives lower-$K$
co-retrieval pairs more influence during partitioning and intra-segment
ordering, prioritizing the more latency-sensitive small-$K$ cases while still
retaining signal from larger retrieval sets.

\noindent\textbf{ANN engines.} Our ANN engines are DiskANN\cite{subramanya2019diskann} and
SPANN\cite{chen2021spann}.
DiskANN is a graph-based SSD ANN engine: compact vector summaries stay in
memory, while graph and data pages are served from SSD; we set the search width
$L$ per $K$ (68--1600 for $K{=}10$--1600) for recall@$K{\approx}0.90$ 
which is a common recall target in evaluation practice of
related work~\cite{subramanya2019diskann,chen2021spann,guo2025pipeann,
guo2026odinann,huang2026helmsman,fusionanns2025}.
For SPANN we set \texttt{SearchInternalResultNum} per $K$ (110--1152 for
$K{=}10$--1600) for recall@$K{\approx}0.90$ and issue posting-list reads in
batches of 8.
Together, these two
engines cover graph-based and cluster-based SSD retrieval, and \sys uses their
returned document IDs plus engine-specific candidate hooks for \sicp
(\S\ref{sec:ssip-impl}). 
We select these representative backends to evaluate both index families,
rather than compare all search-engine variants. PipeANN optimizes
DiskANN-style graph traversal, while OdinANN extends this line to maintain
stable performance under updates~\cite{guo2025pipeann,guo2026odinann};
we do not evaluate them here. Faster vector search makes unchanged content
fetch a larger fraction of retrieval latency, strengthening the motivation
to optimize it, but can also shorten \sicp's overlap window. 
We bounded bandwidth of vector search of SPANN to simulate production 
concurrent environments.

\noindent\textbf{Baselines.} We compare against three baselines.
(1)~Seq\_\allowbreak{}Random stores content in a 
single binary file under a random
corpus insertion order and fetches results via per-document I/O request.
Insertion-order payload storage follows current production practice
\cite{qdrant2024gridstore,wang2021milvus}.
(2)~Seq\_\allowbreak{}KMeans stores content in \ccs segments organized by
embedding K-Means, testing semantic co-location without co-retrieval-aware
grouping.
(3)~Seq\_\allowbreak{}RPAG stores document content in \ccs segments organized
by RPAG and uses the same \ccf-batched fetch path as \sys, but fetches content
only after ANN search completes. 
SICP-only adds speculative prefetch to the Random layout; \sys combines
prefetch with RPAG and \ccf. These variants separate overlap from layout
benefits while preserving the final top-$K$.

\subsection{End-to-End Retrieval Latency}
\label{sec:eval-e2e}

Figures~\ref{fig:e2e-diskann-stacked} and~\ref{fig:e2e-spann-stacked}
show that most savings come from reducing or overlapping foreground content
fetch, while search time changes little. The callback only publishes
candidate snapshots; stability checks and page selection run on the
background worker, limiting search-side overhead.

Seq\_\allowbreak{}RPAG isolates the layout benefit: relative to
Seq\_\allowbreak{}KMeans, it reduces latency by up to 18\% on DiskANN
and 9\% on SPANN, although a few SPANN settings regress by up to 2\%.
The DiskANN benefit peaks near the largest trained prefix and declines
beyond it. This is consistent with RPAG concentrating locality on the
co-retrieval patterns represented in its training objective, rather than
embedding proximity alone.

\sys adds \sicp to hide part of the remaining fetch behind search.
Across per-dataset settings, it reduces end-to-end latency over the best
per-document-read baseline by 4--48\% on DiskANN and up to 25\% on SPANN.
Benefits generally grow with $K$ as more content I/O becomes available to
overlap; small-$K$ SPANN cases show little benefit.
The SICP-only ablation confirms that prefetching benefits even the Random
layout. On DiskANN, layout alone is faster at smaller $K$, while prefetch
alone becomes more effective at large $K$. Combining both gives the lowest
geometric-mean latency at every evaluated $K$ on both engines: grouping
reduces the work, and speculation hides part of what remains.

\subsection{Prefetch Precision and Coverage}
\label{sec:eval-sicp}
We evaluate \sicp's prefetch quality on the \rpag\ layout used by \sys in
Figures~\ref{fig:e2e-diskann-stacked} and~\ref{fig:e2e-spann-stacked}.
\begin{figure}[t!]
\centering
\includegraphics[width=1.9in]{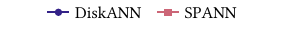}\par
\vspace{1pt}
\begin{subfigure}[t]{0.49\columnwidth}
  \centering
  \includegraphics[width=\linewidth]{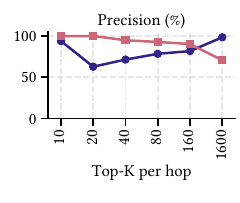}
  \caption{Precision}
  \label{fig:prefetch-precision}
\end{subfigure}\hfill
\begin{subfigure}[t]{0.49\columnwidth}
  \centering
  \includegraphics[width=\linewidth]{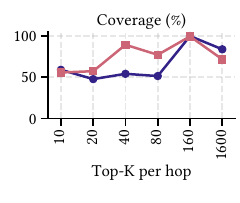}
  \caption{Coverage}
  \label{fig:prefetch-coverage}
\end{subfigure}
\caption{Prefetch precision and final-K coverage}
\label{fig:prefetch-quality}
\end{figure}

Figure~\ref{fig:prefetch-quality} shows that many selected candidates survive
into the final results: precision is 62.6--98.3\% for DiskANN and
70.4--99.9\% for SPANN. Both cover over 99\% of results at $K{=}160$.

\begin{figure}[t]
\centering
\includegraphics[width=2.6in]{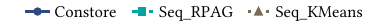}\par
\vspace{1pt}
\begin{subfigure}[t]{0.49\columnwidth}
  \centering
  \includegraphics[width=\linewidth]{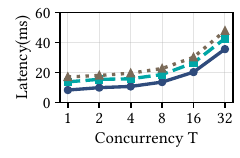}
  \caption{SPANN}
  \label{fig:concurrency-thread-spann}
\end{subfigure}\hfill
\begin{subfigure}[t]{0.49\columnwidth}
  \centering
  \includegraphics[width=\linewidth]{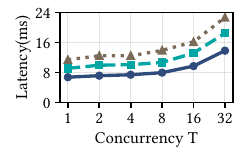}
  \caption{DiskANN}
  \label{fig:concurrency-thread-diskann}
\end{subfigure}
\caption{End-to-end latency as query concurrency increases. }
\label{fig:concurrency-thread-latency}
\end{figure}

\subsection{Scalability}
\label{sec:eval-concurrency}
Figure~\ref{fig:concurrency-thread-latency} compares \sys with
Seq\_\allowbreak{}KMeans and Seq\_\allowbreak{}RPAG as concurrency grows
from one to 32 queries. \sys retains the lowest latency on both engines.
Its single search thread per query only publishes snapshots; the worker
overlaps bounded content reads with search. RPAG and \ccf reduce the
content-I/O demand that would otherwise compete with concurrent searches.

\begin{figure}[t!]
\centering

\includegraphics[width=2.60in]{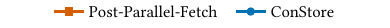}\par
\vspace{1pt}
\begin{subfigure}[t]{0.49\columnwidth}
  \centering
  \includegraphics[width=\linewidth]{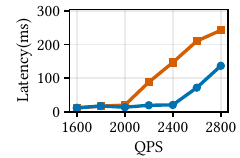}
  \caption{DiskANN}
\end{subfigure}\hfill
\begin{subfigure}[t]{0.49\columnwidth}
  \centering
  \includegraphics[width=\linewidth]{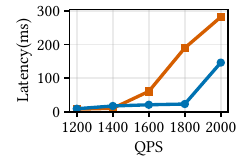}
  \caption{SPANN}
\end{subfigure}
\caption{Latency vs. offered QPS }
\label{fig:concurrency-qps-p99}

\vspace{0.6em}

\includegraphics[width=2.60in]{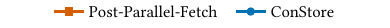}\par
\vspace{1pt}
\begin{subfigure}[t]{0.49\columnwidth}
  \centering
  \includegraphics[width=\linewidth]{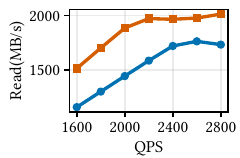}
  \caption{DiskANN}
\end{subfigure}\hfill
\begin{subfigure}[t]{0.49\columnwidth}
  \centering
  \includegraphics[width=\linewidth]{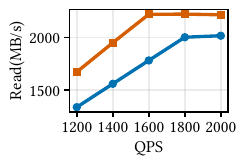}
  \caption{SPANN}
\end{subfigure}
\caption{Device read bandwidth vs. offered QPS }
\label{fig:concurrency-resource-qps}

\end{figure}

Figure~\ref{fig:concurrency-qps-p99} compares \sys with
Post-Parallel-Fetch, which parallelizes content reads only after search.
Under a 100\,ms total-P99 SLO and a 95\% achieved/offered-QPS gate,
\sys increases sustainable offered QPS by 20\% on DiskANN and 12.5\% on
SPANN. At the baselines' highest sustainable rates 
, P99 latency is lower by factors of $1.46\times$ and
$2.90\times$.

Both engines therefore sustain higher load before saturation: bounded
search--fetch overlap reduces exposed fetch time without enough added
contention to erase the benefit.

Device statistics show that this improvement does not require more read
bandwidth (Figure~\ref{fig:concurrency-resource-qps}). At those same
operating points, \sys lowers device read bandwidth by 23.4\% on DiskANN
and 19.7\% on SPANN. Engine-accounted content traffic and I/O requests also
fall, consistent with grouped fetch reducing pressure on search-phase I/O.

\begin{figure*}[t]
\centering
\includegraphics{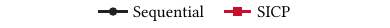}\par
\vspace{1pt}
\begin{subfigure}[t]{0.48\textwidth}
  \centering
  \includegraphics[width=\linewidth]{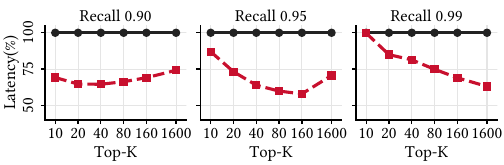}
  \caption{DiskANN, SIFT100M}
  \label{fig:sift-diskann-100m}
\end{subfigure}
\hfill
\begin{subfigure}[t]{0.48\textwidth}
  \centering
  \includegraphics[width=\linewidth]{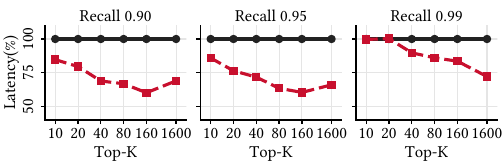}
  \caption{DiskANN, SIFT1B}
  \label{fig:sift-diskann-1b}
\end{subfigure}

\begin{subfigure}[t]{0.48\textwidth}
  \centering
  \includegraphics[width=\linewidth]{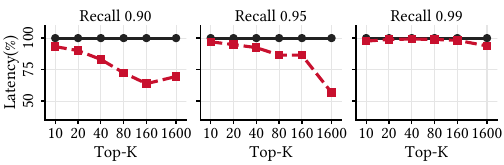}
  \caption{SPANN, SIFT100M}
  \label{fig:sift-spann-100m}
\end{subfigure}
\hfill
\begin{subfigure}[t]{0.48\textwidth}
  \centering
  \includegraphics[width=\linewidth]{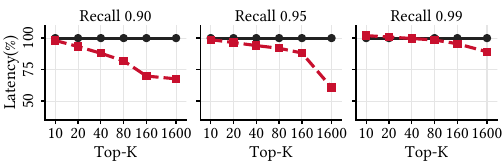}
  \caption{SPANN, SIFT1B}
  \label{fig:sift-spann-1b}
\end{subfigure}
\caption{End-to-end latency of \sicp relative to Sequential across top-$K$ and
recall targets.}
\label{fig:sift-diskann-scaling}
\end{figure*}

\begin{figure*}[!t]
\centering
\begin{minipage}[t]{\columnwidth}
\vspace{0pt}
\centering
\includegraphics[width=\linewidth]{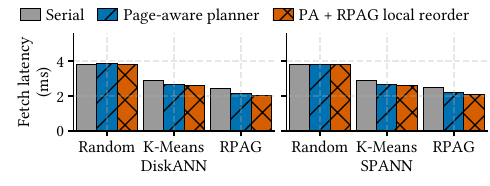}
\caption{Content-fetch latency by layout and fetch engine}
\label{fig:fetch-p50}
\end{minipage}\hfill
\begin{minipage}[t]{\columnwidth}
\vspace{0pt}
\centering
\captionof{table}{Cross-round segment hit rate (\%).}
\label{tab:cross-round}
\begin{tabular}{lrr}
\toprule
Layout & DiskANN & SPANN \\
\midrule
Random & 3.9 & 3.8 \\
K-Means & 15.5 & 15.4 \\
RPAG & 52.8 & 49.6 \\
\bottomrule
\end{tabular}
\end{minipage}
\end{figure*}

\subsection{Scaling with Top-$K$ and Index Size}
\label{sec:eval-scale-k}
We evaluate \sicp on SIFT100M and SIFT1B at recall targets 0.90, 0.95,
and 0.99. Sequential fetches the synthetic payloads after search; \sicp
overlaps reads from the same insertion-order file. 

Figure~\ref{fig:sift-diskann-scaling} shows that overlap remains useful at
both scales without changing the final top-$K$. At recall targets
0.90/0.95, DiskANN's latency reductions reach 42\% on SIFT100M and 40\%
on SIFT1B; SPANN saves 30--43\% at $K{=}1600$ across the two scales.
The gain depends on the search/fetch balance, not scale alone: at SPANN's
0.99 target, search dominates and the corresponding savings fall to 6--11\%.
Thus, larger result sets create more work to overlap, but a search-dominated
pipeline limits the relative end-to-end benefit.

\subsection{Content Fetch Speedup}
\label{sec:eval-layout}
We now turn to the second technique and isolate the layout side, measuring
content fetch after search completes with no prefetch involved.

\noindent\textbf{Page-aware fetch and local ordering.}
Figure~\ref{fig:fetch-p50} compares Serial (no page-aware coalescing),
Page-aware planner (same layout, aligned coalesced reads), and
PA + RPAG local reorder (intra-segment reordering followed by the planner).
Planning helps both K-Means and RPAG; reordering further places
co-retrieved documents on nearby pages, making each read more useful.
With both optimizations, RPAG lowers fetch latency by 45--47\% over Random
and 19--22\% over K-Means using the same optimized fetch path.
The advantage therefore comes from co-retrieval locality as well as the
shared planner.

\noindent\textbf{Cross-hop reuse opportunity.}
RPAG places about half of hop-2 results in segments touched by hop-1,
versus roughly 15\% for K-Means and 4\% for Random
(Table~\ref{tab:cross-round}). Grouping co-retrieved evidence thus creates
reuse opportunities across hops on both engines. This is a layout-derived
upper bound, not a measured cache-hit rate: hop-1 may read only some pages
of a segment, and actual reuse requires those pages to remain cached.

\subsection{Layout Staleness and Adaptation}
\label{sec:eval-wasr}

We replay the DiskANN hop traces as a chronological stream of 500-query
epochs under a zipfian query popularity. Four warm-up epochs build the
initial layout; each later epoch is evaluated before its queries reach the
monitor. For example, ``WASR 0.5\%'' adds rewrite allowance equal to
0.5\% of corpus bytes each epoch. Unused allowance accumulates in a capped
token bucket, so a single reorganization can exceed one epoch's allowance;
rewrites occur only when the adaptation checks in \S\ref{sec:wasr} pass.
Benefit retention is $(L_{\text{kmeans}}-L)/(L_{\text{kmeans}}-
L_{\text{oracle}})$, where $L$ is the mean number of segments touched per
two-hop query under the evaluated policy, and $L_{\text{kmeans}}$ and
$L_{\text{oracle}}$ are the corresponding means for K-Means and the oracle.
These quantities measure segment accesses, not latency. The oracle is a
full rebuild that knows the confirmed graph of the next four epochs.

\begin{figure}[t]
\centering
\includegraphics[width=\columnwidth]{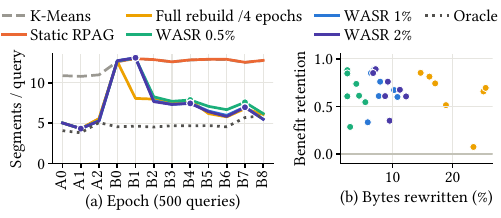}
\caption{(a) Segments touched per two-hop query across
epochs; the workload switches  at epoch B0. (b)
Benefit retained relative to an oracle rebuild vs. bytes rewritten.
}

\label{fig:wasr}
\end{figure}

Figure~\ref{fig:wasr}(a) shows that static \rpag\ loses its locality
advantage after the workload shift. \wasr\ detects the degradation and
restores locality through selective reorganization: with a 1\% per-epoch
budget, it rewrites 10.8\% of the corpus in total, versus 25\% for periodic
full rebuilds. Targeting affected segments thus adapts the layout with
substantially less rewriting.

Across the seven scenarios in Figure~\ref{fig:wasr}(b), \wasr\ rewrites
2.5--12\% of the corpus, compared with 15--26\% for periodic full rebuilds.
Its recurrence and gain checks focus this budget on shifts with reusable
co-retrieval patterns rather than rebuilding on a fixed schedule. On the
unique-query replay, \wasr\ avoids rewriting entirely because there is
insufficient recurring evidence, while periodic rebuilds still incur
rewrite cost.

\section{Related Work}
\label{sec:related}

\noindent\textbf{Vector search optimization.} A long line of ANN work reduces the cost of
candidate generation through compression, graph navigation, filtering, and
specialized hardware execution. FAISS, product quantization, optimized product
quantization, ScaNN, HNSW, and NSG improve vector representation or in-memory
search efficiency~\cite{johnson2019faiss,jegou2011pq,ge2013opq,guo2020scann,
malkov2020hnsw,fu2019nsg}. DiskANN and SPANN move billion-scale search to SSDs;
Filtered-DiskANN, FreshDiskANN, SPFresh, Starling, and FusionANNS extend this
line with filters, updates, disk-layout optimization, or CPU/GPU
execution~\cite{subramanya2019diskann,chen2021spann,gollapudi2023filtereddiskann,
singh2021freshdiskann,xu2023spfresh,wu2024starling,fusionanns2025}. ANN
benchmarking work characterizes the recall/latency tradeoff of such
indexes~\cite{aumuller2020annbench}. These systems optimize Phase~1
search I/O or scoring; they do not optimize Phase~2 content fetch.

\noindent\textbf{Retrieval and RAG models.} Neural retrieval systems such as REALM,
DPR, RAG, FiD, RETRO, and Atlas show that generation quality depends on retrieving
external evidence, often at large $K$ or over very large corpora~\cite{guu2020realm,
karpukhin2020dpr,lewis2020rag,izacard2021fid,borgeaud2022retro,
izacard2023atlas}. More recent active, self-reflective, corrective, and
interleaved retrieval methods make retrieval part of the generation loop
\cite{jiang2023flare,asai2024selfrag,yan2024crag,trivedi2023ircot,
gao2024ragsurvey}. Dense and late-interaction retrievers such as ColBERT,
ColBERTv2, and PLAID further improve retrieval quality and efficiency, while
BEIR and Natural Questions broaden retrieval evaluation~\cite{khattab2020colbert,
santhanam2022colbertv2,santhanam2022plaid,thakur2021beir,kwiatkowski2019nq}.
These works motivate larger and repeated evidence fetches, but they do not
change how retrieved payloads are placed on storage.

\noindent\textbf{RAG and LLM serving systems.} RAGCache, Mooncake, and RAGO optimize
RAG serving through KV-cache reuse, disaggregated KV-cache management, or
pipeline scheduling~\cite{ragcache2024,mooncake2025,rago2025}. General LLM
serving systems such as Orca, vLLM, SGLang, FlexGen, DistServe, Sarathi-Serve,
Splitwise, and ZeRO-Infinity improve batching, memory management, offloading, or
phase disaggregation for model execution~\cite{yu2022orca,kwon2023vllm,
zheng2024sglang,sheng2023flexgen,zhong2024distserve,agrawal2024sarathi,
patel2024splitwise,rajbhandari2021zeroinfinity}. They optimize the LLM
side of the pipeline; \sys optimizes the storage-side content path that feeds the
prompt.

\noindent\textbf{Storage for AI workloads.} Recent AI-storage systems tailor data
formats and runtime scheduling to ML workloads. ServerlessLLM introduces
loading-optimized checkpoint formats, FlacIO reduces I/O amplification for ML
artifacts, and CacheLib provides a production DRAM/SSD caching engine for
object workloads~\cite{serverlessllm2024,flacio2025,berg2020cachelib,dumpkv}. These
systems share \sys's workload-aware philosophy, but they target model loading,
container or artifact I/O, and caching rather than the co-retrieval layout of RAG
content.

\noindent\textbf{Vector DB payload storage.} Classic storage systems such as LFS,
LSM-trees, RocksDB, and WiscKey optimize write-optimized layout, compaction, or
key-value separation for general storage workloads~\cite{rosenblum1992lfs,
oneil1996lsm,dong2021rocksdb,lu2016wisckey}. Production vector databases adopt
practical payload stores: Milvus and MANU use binlog/object-store style layouts,
Qdrant replaced RocksDB with Gridstore, and Weaviate uses a custom LSM
\cite{wang2021milvus,guo2022manu,qdrant2024gridstore,weaviate2024lsmkv}.
These systems store payloads by insertion order, key order, or segment
format; none use co-retrieval-aware layout or implement GCF-style batching.

\section{Conclusion}
\label{sec:conclusion}

\sys treats content fetch as a first-class cost of disk-based ANNS retrieval.
\sicp overlaps bounded content reads with search, while \rpag and \ccf
reduce fetch I/O through co-retrieval-aware layout. The techniques are
complementary and preserve the final top-$K$; together, they reduce
end-to-end retrieval latency by up to 48\% on DiskANN and 25\% on SPANN
over the best per-document-read baseline.

\end{document}